\documentclass[10pt,conference]{IEEEtran} 
\IEEEoverridecommandlockouts
\usepackage{cite}
\usepackage{amsmath,amssymb,amsfonts}
\usepackage{algorithmic}
\usepackage{graphicx}
\usepackage{textcomp}
\usepackage{xcolor}
\usepackage{hyperref}
\usepackage{longtable}
\usepackage{multirow}
\usepackage{url} 
\usepackage{array}
\usepackage{booktabs}
\usepackage{colortbl}
\usepackage{tcolorbox}

\def\BibTeX{{\rm B\kern-.05em{\sc i\kern-.025em b}\kern-.08em
    T\kern-.1667em\lower.7ex\hbox{E}\kern-.125emX}}

\newcommand{\hierarchical}{\text{hierarchical}}
\newcommand{\flatp}{\text{flat}}

\begin{document}
\title{Exploring the Potential of Large Language Models in Fine-Grained Review Comment Classification}

\author{
  \IEEEauthorblockN{Linh Nguyen, Chunhua Liu, Hong Yi Lin, Patanamon Thongtanunam}
  \IEEEauthorblockA{\textit{The University of Melbourne} \\
  \textit{Melbourne, Australia} \\
  linh.nguyen3@student.unimelb.edu.au\\
  \{chunhua.liu1, tom.lin1, patanamon.t\}@unimelb.edu.au}
}

\maketitle

\begin{abstract}
Code review is a crucial practice in software development. As code review nowadays is lightweight, various issues can be identified, and sometimes, they can be trivial. Research has investigated automated approaches to classify review comments to gauge the effectiveness of code reviews. However, previous studies have primarily relied on supervised machine learning, which requires extensive manual annotation to train the models effectively. To address this limitation, we explore the potential of using Large Language Models (LLMs) to classify code review comments. We assess the performance of LLMs to classify 17 categories of code review comments. Our results show that LLMs can classify code review comments, outperforming the state-of-the-art approach using a trained deep learning model. In particular, LLMs achieve better accuracy in classifying the five most useful categories, which the state-of-the-art approach struggles with due to low training examples. Rather than relying solely on a specific small training data distribution, our results show that LLMs provide balanced performance across high- and low-frequency categories.
These results suggest that the LLMs could offer a scalable solution for code review analytics to improve the effectiveness of the code review process.

\end{abstract}


\section{Introduction}
Code Review (CR) is a practice in software development where developers review other developer's code changes asynchronously to find defects and suggest improvements \cite{Bacchelli}. Acting as a quality assurance gateway, CR has become mandatory in many prominent organizations, with developers reportedly spending 10-15\% of their time on this task \cite{Bosu2016}. 
In practice, various types of concerns can be raised in CR comments, ranging from code styling to functional issues.
As comments often trigger the improvements of code changes, the types of comments play a crucial role in the quality of CR.

Constructive and actionable comments addressing quality-improving issues would positively contribute to CR's overall quality and code changes~\cite{Bird,Czerwonka,thongtanunam2015investigating}.
On the other hand, trivial or irrelevant comments can waste developers' time without improving the code changes~\cite{Bosu2015}. 

For example, a comment flagging a deprecated Python method indicates a functional defect requiring immediate attention to prevent potential issues, whereas a suggestion for visual representation is generally less urgent.
Indeed, due to the lightweight variant of modern code reviews, prior studies indicate that 20\% to 44\% of comments are considered ``not useful" by authors, underscoring an urgent need for enhanced CR effectiveness \cite{Bosu2015, hasan, Rahman, TurzoBosu}.
Automating the classification of such comments provides insights into the quality of current CR practices, identifies areas for improvement, and enhances their effectiveness.

Prior research has investigated an automated approach to classify code review comments~\cite{pangsakulyanont2014assessing,Bosu2015,ebert2017confusion,Yang-etal-2023-EvaCRC, Fregnan-etal-2022-classifying, TurzoBosu_new}.
Recently, Turzo et al.~\cite{TurzoBosu_new} leveraged deep learning techniques to classify comments into five high-level groups, e.g., function issues, and refactoring. 
Specifically, they trained a deep learning model (CodeBERT and LSTM) using 1,828 manually annotated comments from their previous work~\cite{TurzoBosu}.
While this approach shows promise for analyzing comment semantics, supervised learning methods require extensive manual effort to prepare training data, which can limit their generalizability.

Instead of the five high-level groups, review comments can be classified into 17 fine-grained categories that more accurately capture the concerns behind review comments~\cite{Mika}. 
A recent study also has established a link to the usefulness rated by practitioners~\cite{TurzoBosu}.
Their results also highlight that the five most useful types are scarce within CR.
For instance, functional defects—the most useful type—make up less than 1\% of the CR comments.
Despite its importance, it's challenging to classify comments at this fine-grained level due to the limited and imbalanced data.

In this work, we explore the capability of Large Language Models (LLMs) to analyze and classify code review comments.
LLMs have demonstrated substantial potential across various coding and natural language processing tasks~\cite{Chang-etal-2024-surveyLLM,Zhao-etal-2023-LLMSurvey, zan-etal-2023-large}.
As LLMs can perform tasks directly without further training or fine-tuning on specific data, LLMs’ pre-trained code knowledge and broad understanding of linguistic patterns make them suitable for handling imbalanced data like the review comment categories~\cite{Beller, Mika}.
This lightweight setup could enhance practical adoption.

Despite the promising capability, LLMs are sensitive to query framing and context length, leading to varied responses~\cite{nam2024,anagnostidis2024}. 
Thus, we set out to explore different types of information (e.g., definitions, code context) and classification strategies (e.g., single or two-step) across different LLM sizes.
We compare the LLM approaches with the state-of-the-art approach~\cite{TurzoBosu_new}, which trained a small deep learning model from scratch.
We address the following research questions:

\begin{enumerate} 

\item[\textbf{RQ1:}] \textbf{Can we use  LLMs to classify code review comments? }
With the large model (Llama 3.1-405B), using the single-step classification with code context achieves the highest average F1-score of 46.2. 
For the small and medium models (e.g., Llama 3-8B, Qwen 2-72B), 
using the two-step classification could help boost their performance.
These results suggest that LLMs have the potential to classify review comments into 17 categories, despite the highly uneven distribution of frequencies.

\item[\textbf{RQ2:}] \textbf{Do LLMs outperform  the state-of-the-art approach?}
LLMs can outperform the state-of-the-art approach which trained a small deep learning model from scratch~\cite{TurzoBosu}. 
When using Llama 3.1-405B with the single-step classification, the F1 score is improved by 10\% and 11.3\% with and without code, respectively. Nonetheless, using the smaller models (e.g., Llama 3-70B or Qwen 2-72B) will result in a lower performance than such a supervised deep learning approach.

\item[\textbf{RQ3:}] \textbf{Which categories can LLMs accurately classify?} \\
We found that LLMs are effective at classifying praise and visual representation with an F1 score of 62.6 and 71.4, respectively.
Moreover, LLMs substantially outperform the state-of-the-art method in the five most
useful categories, i.e., functional defects, validation, logical, interface, and solution approach. Additionally, our approach shows a
considerable improvement in categories where the state--
of-the-art method struggles due to insufficient data.

\end{enumerate}

\textbf{Significance \& Contribution.}
Our results highlight the advantages of leveraging LLMs’ pre-trained knowledge through prompt engineering, which mitigates challenges associated with limited labeled data in code review classification—an issue that can impact such a supervised deep learning model like the state-of-the-art approach. 
Our LLM-based approach demonstrated more balanced performance across both high- and low-frequency categories, counteracting the class imbalance issues, making these LLMs particularly effective for datasets with uneven category distributions.
Our LLM-based approaches could offer a scalable solution for code review analytics to improve the effectiveness of the CR process.

\textbf{Novelty.} To the best of our knowledge, this paper is the first to 1) present an automated approach to classify review comments into 17 categories; and 2) empirically investigate the capability of LLMs for review comment classification; and 3) analyze the strengths and limitations of LLMs on fine-grained code review categories.

\textbf{Open Science.} To facilitate future work, we provide a replication package containing scripts for both LLM and state-of-the-art approaches, as well as experimental results.~\footnote{\url{https://zenodo.org/records/15003074}}

\textbf{Paper Organisation.}
The remainder of this paper is structured as follows: 
Section~\ref{sec:background} discusses related work. 
Sections~\ref{sec:design} and \ref{sec:exp_setup} outline an overview of the study design and experimental setup. 
Sections~\ref{sec:results} - \ref{sec:discussion} present and discuss the results.  Section~\ref{sec:threat} discusses potential threats. Finally, Section~\ref{sec:conclusion} draws a conclusion.

\section{Background and Related Works}
\label{sec:background}
\subsection{The Importance of Code Review Comments}
The code review (CR) process is an essential practice in software development, helping teams identify defects, improve code quality, and foster knowledge sharing \cite{Bosu2015}. The effectiveness of CR largely depends on the nature of review comments~\cite{thongtanunam2015investigating,TurzoBosu}.
Constructive and relevant feedback can enhance code quality, while misguided or trivial comments risk wasting developers' time without contributing meaningfully to improvements \cite{Bird}. 
Studies reveal a mismatch between the intended benefits of CRs and actual outcomes~\cite{Czerwonka}, for example, while managers and developers hope to uncover significant issues, CRs often reveal only minor, localized defects rather than critical issues \cite{Beller}. 
This discrepancy expectation highlights the need for CR analytics to evaluate and reflect the effectiveness of CR, which impacts code quality.
Displaying information about the code review process can encourage teams to reflect on their practices\cite{Bird,Fregnan-etal-2022-classifying}. For example, if most reviews focus on visual issues, these could be addressed earlier using static analysis tools or improved style guidelines. This would free up developers to focus on critical functional defects.


\subsection{Code Review Comment Types \& Their Usefulness}
To assess the effectiveness of CR, several studies examined the review comment types in modern code reviews.
Using the defect-based taxonomy~\cite{Mika}, Beller et al. found that 75\% of the CR are related to maintainability and 25\% are related to functional issues.
Thongtanunam et al.~\cite{thongtanunam2015investigating} found that defective files tend to undergo review that often discusses maintainability rather than functional issues.
Recent studies also reveal that reviewers sometimes express confusion in CR comments~\cite{ebert2021exploratory} or even anti-patterns like toxic discussion or identifying trivial issues~\cite{chouchen2021anti}.

To better understand the importance of each type of CR comment, Turzo and Bosu.~\cite{TurzoBosu} conducted a survey study with software practitioners to rank the usefulness of CR comments. 
Table \ref{tab:review_comment_taxonomy} provides an overview of review comment types and their rank of usefulness.
They found that comments related to functional defects, validation, logic, interface, and solution approach are the five most useful comments.
Other comments like documentation, code organization, and visual representation are considered less useful.

\begin{table*}
    \centering
    \caption{Review comment types, their usefulness rating and frequency~\cite{TurzoBosu}.}
\begin{tabular}{l|l|p{10cm}|c|c} 
\toprule
\textbf{Group} & \textbf{Category} & \textbf{Description}  & \textbf{Rating$\dag$}  & \textbf{Frequency}  \\ 
\midrule
 
 \parbox[t]{2mm}{\multirow{9}{*}{\rotatebox[origin=c]{90}{\textbf{Functional}}}} 
 & Functional defects$^*$ & Functionality is missing or implemented incorrectly and such defects often require additional code or larger modifications to the existing solution. & 4.38 & 12 (0.65\%)\\ \cline{2-5}

& Logical$^*$  & Control flow, comparison related, and logical errors. & 4.11 & 56 (3\%)\\ \cline{2-5}

& Validation$^*$  & Validation mistakes or mistakes made when detecting an invalid value are of this class. Any kind of user data sanitization-related comments are in this category, too. & 4.16 & 90 (4.92\%)\\ \cline{2-5}

&  Resource  & Resource (variables, memory, files, database) initialization, manipulation, and release. & 3.83 & 34 (1.85\%)\\ \cline{2-5}

& Timing & Potential issues due to incorrect thread synchronization. &  3.5 & 4 (0.21\%)\\ \cline{2-5}

& Support issues & Issues related to support systems and libraries or their configurations. & 3.51 & 14 (0.76\%) \\ \cline{2-5}

 & Interface$^*$  & Mistakes when interacting with other parts of the software such as existing code library, hardware device, database, or operating system. & 4.10 & 30 (1.6\%) \\ \hline

 \parbox[t]{4mm}{\multirow{5}{*}{\rotatebox[origin=c]{90}{\textbf{Refactoring}}}} 
& Solution approach$^*$ & Suggestions to adopt an alternate algorithm or data structure. & 4.00 & 201 (11\%)\\ \cline{2-5}

 & Code Organization  & Refactoring suggestions such as those included in Martin Fowlers’s catalog. & 3.68 & 184 (10\%)\\ \cline{2-5}

 & Alternate Output  & Comments that suggest modifying the error message, toast message, alert, or change what is returned by a function. & 3.63 & 64 (3.5\%)\\  \cline{2-5}

& Naming Convention & Violations of identifier naming conventions. & 3.43 & 76 (4.15\%) \\ \cline{2-5}

& Visual Representation & Whitespace, blank lines, code rearrangements, and indentation-related comments. & 2.92 & 73 (4\%) \\  \hline

 \multicolumn{2}{l|}{\textbf{Documentation}}  & Suggestions to add /modify comments or documentation to aid code comprehension. & 3.73 & 387 (21\%) \\ 
\hline

 \parbox[t]{2mm}{\multirow{3}{*}{\rotatebox[origin=c]{90}{\textbf{Discuss.}}}} 
& Question & Questions to understand the design or implementation choices. & 3.99 & 275 (15\%)  \\ \cline{2-5}

& Design discussion & Discussions on design direction, design pattern, and software architecture. & 3.87 & 87 (4.75\%) \\ \cline{2-5} 

& Praise & Complement for a code. & 3.03 & 83 (4.5\%) \\ \hline

 \multicolumn{2}{l|}{\textbf{False positive}} & If a review comment raises an invalid bug or concern. & N/A & 158 (8.6\%) \\ 
\bottomrule
\multicolumn{5}{l}{\footnotesize $\dag$ An average rating based on the practitioner survey~\cite{TurzoBosu}, where a score of 5 indicates ``very useful'' and a score of 1 indicates ``least useful''. } \\
\multicolumn{5}{l}{\footnotesize $*$ Five most useful types of review comments ranked by practitioners. }
\end{tabular}
\label{tab:review_comment_taxonomy}
\end{table*}




\subsection{Automated Review Comment Classification}
To gauge overall effectiveness and gain insight into CR practices, several studies explored automated approaches for review comment classification. 
Previous work has primarily focused on three classification targets: 1) usefulness, 2) linguistic patterns, and 3) defect types.

For classifying usefulness, the classification task is formulated as a binary task, i.e., whether a review comment is useful or not.
For example, Pangsakulyanont et al.~\cite{pangsakulyanont2014assessing} proposed that the usefulness can be measured based on semantic similarity between a review comment and code changes.
Bosu et al.~\cite{Bosu2015} developed a decision tree based on various signals within the CR process to identify useful review comments. 
While this classification task can help practitioners assess overall code review effectiveness (e.g., measuring the proportion of useful comments), it provides limited insights into the comments themselves.

Rather than focusing on usefulness, studies focus on linguistic
patterns.
For example, Ebert et al.~\cite{ebert2017confusion} trained a machine-learning model to classify comment that expresses confusion.
Recently, Yang et al.~\cite{Yang-etal-2023-EvaCRC} leveraged a deep learning model to classify a comment into four types: emotion, question, evaluation, and suggestion, which captures communication styles.
Although this classification provides insights into communication dynamics and reviewer intent, it did not capture technical concerns of code changes under review.

Inspired by empirical studies~\cite{Beller,Mika}, research has explored approaches to classify code review comments into the defect-based taxonomy. 
Initially, Fregnan et al.~\cite{Fregnan-etal-2022-classifying} trained machine learning classifiers (Naive Bayes, Decision Tree, and Random Forest) using software and process features.
Turzo et al.~\cite{TurzoBosu_new} explored a deep learning model to analyze the semantics of the review comments.
In particular, they used CodeBERT--a pre-trained language model to generate a representation of a code review comment then trained an LSTM model to classify comments into five high-level groups (see Table \ref{tab:review_comment_taxonomy}). 
They have outperformed feature-based approaches, establishing themselves as the state-of-the-art method.
While the state-of-the-art classification approach captures valuable insights into code review comments, it relies on labeled data for training, which is not scalable and may limit its generalizability to other datasets or programming languages. 

As an empirical study~\cite{TurzoBosu} has established a relationship between the 17 review comment types and the usefulness rating, a more granular classification of review comments into 17 types could offer clearer insights analytics on code review effectiveness. 
However, such a fine-grained classification is not trivial, given the nuanced nature of the comments and the imbalanced data.
As shown in Table~\ref{tab:review_comment_taxonomy}, the dataset is imbalanced.
The Documentation and Question categories are majority, accounting for 21\%  and  15\% of the comments, respectively.
On the other hand, the more useful categories such as functional defect and interface have a very small proportion (under 10\%).
Small deep learning models trained from scratch may struggle with categories that have few examples, resulting in poor performance.

\section{Study Design}
\label{sec:design}
In this work, instead of classifying the five high-level categories, we aim to classify the 17 more specific categories, which are directly linked to practitioners' usefulness ratings~\cite{TurzoBosu}. Since LLMs do not require fine-tuning with specific data, we explore their ability to analyze and classify code review comments, even with a highly unbalanced category distribution.



\subsection{Research Questions}
To empirically evaluate the capabilities of LLMs in the classification of code review comments without additional training, we formulate three research questions.
    
\begin{enumerate}
    \item[\textbf{RQ1:}] \textbf{Can we use  LLMs to classify code review comments? } 
LLMs have shown promise in directly transferring knowledge that is not trained \cite{Brown-etal-2020-few-shot,Hou-etal-2024}.
However, their capability to analyze and classify code review comments is largely unexplored. 
Thus, to address RQ1, we explore prompts with various contextual information and structure to classify review comments.
This will help us identify promising prompt instructions for LLMs in automated review comment classification.

\item[\textbf{RQ2:}] \textbf{Do LLMs outperform the state-of-the-art approach?}
Prior work~\cite{TurzoBosu_new} has trained small deep learning models from scratch to classify code review comments, achieving state-of-the-art performance in classifying five high-level categories and surpassing traditional machine learning methods such as Random Forest.
Yet, little is known whether LLMs which do not require task-specific fine-tuning, can outperform the supervised deep learning approach.
Thus, we conduct a comparative evaluation of LLMs and the state-of-the-art supervised deep learning approach for classifying code review comments.


\item[\textbf{RQ3:}] \textbf{Which categories can LLMs accurately classify?}
Prior work~\cite{TurzoBosu} established a mapping between each category and practitioners' ratings of usefulness.
Yet, it is unclear which categories the LLM-based approach performs well. 
An approach that accurately identifies useful comments would offer a better gauge of overall code review effectiveness. 
To address RQ3, we analyze the effectiveness of LLM and state-of-the-art approaches for each review comment category. 
This insight will help us better understand the potential benefits and areas for improvement.
\end{enumerate}

\subsection{LLM-based Review Comment Classification}
\label{sec:llm_approach}
This section describes our prompting approaches to instruct the LLM to perform a classification task.
We explore two key aspects: 1) classification structure and 2) input context.

\subsubsection{Classification Strategies}
\begin{figure*}
    \centering
    \includegraphics[width=\linewidth]{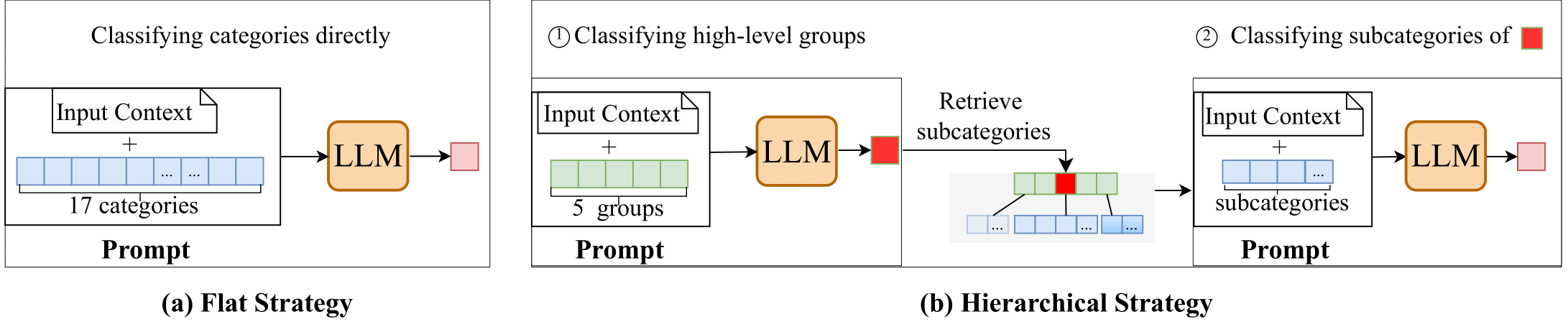}
    \caption{Our classification strategies for classifying review comments are (a) flat (a single-step process) and (b) hierarchical (a two-step process). }
    \label{fig:prompt_structure}
\end{figure*}
Since the review comment taxonomy is organized into two levels: 5 groups and 17 categories (see Table \ref{tab:review_comment_taxonomy}),
we aim to classify review comments aligning with this taxonomy structure.
Therefore, we explore two classification strategies: \textbf{\flatp{}} and \textbf{\hierarchical}.
Figure~\ref{fig:prompt_structure} illustrates our classification strategies with LLMs.


The \textbf{\flatp} strategy is a single-step process that focuses only on the fined-grained level of the taxonomy.
An LLM will be instructed to classify a review comment $R$ into one of the 17 categories. 
This strategy allows an LLM to consider all possible categories for $R$.
However, given long context input (e.g., category definitions), it can be challenging for LLMs to reason across many relations and determine all 17 categories simultaneously.

To avoid a long context input, we also explore a \textbf{\hierarchical} strategy.
It is a two-step process that uses a high-level group to narrow down the categories, reducing the number of options the model needs to consider.
Specifically, the \hierarchical{} strategy will first instruct an LLM to select one of the five high-level groups that is relevant to the review comment $R$.
Then,  the LLM will be further instructed to classify $R$ into one of the categories under the selected high-level group. This decomposition reduces the context input and classification space, reducing task difficulty at each step. However, any errors made in the high-level classification could propagate to the subcategories, which lead to inaccuracies.

\subsubsection{Prompt Design \& Input Context}
\begin{figure}
    \centering
    \includegraphics[width=\linewidth]{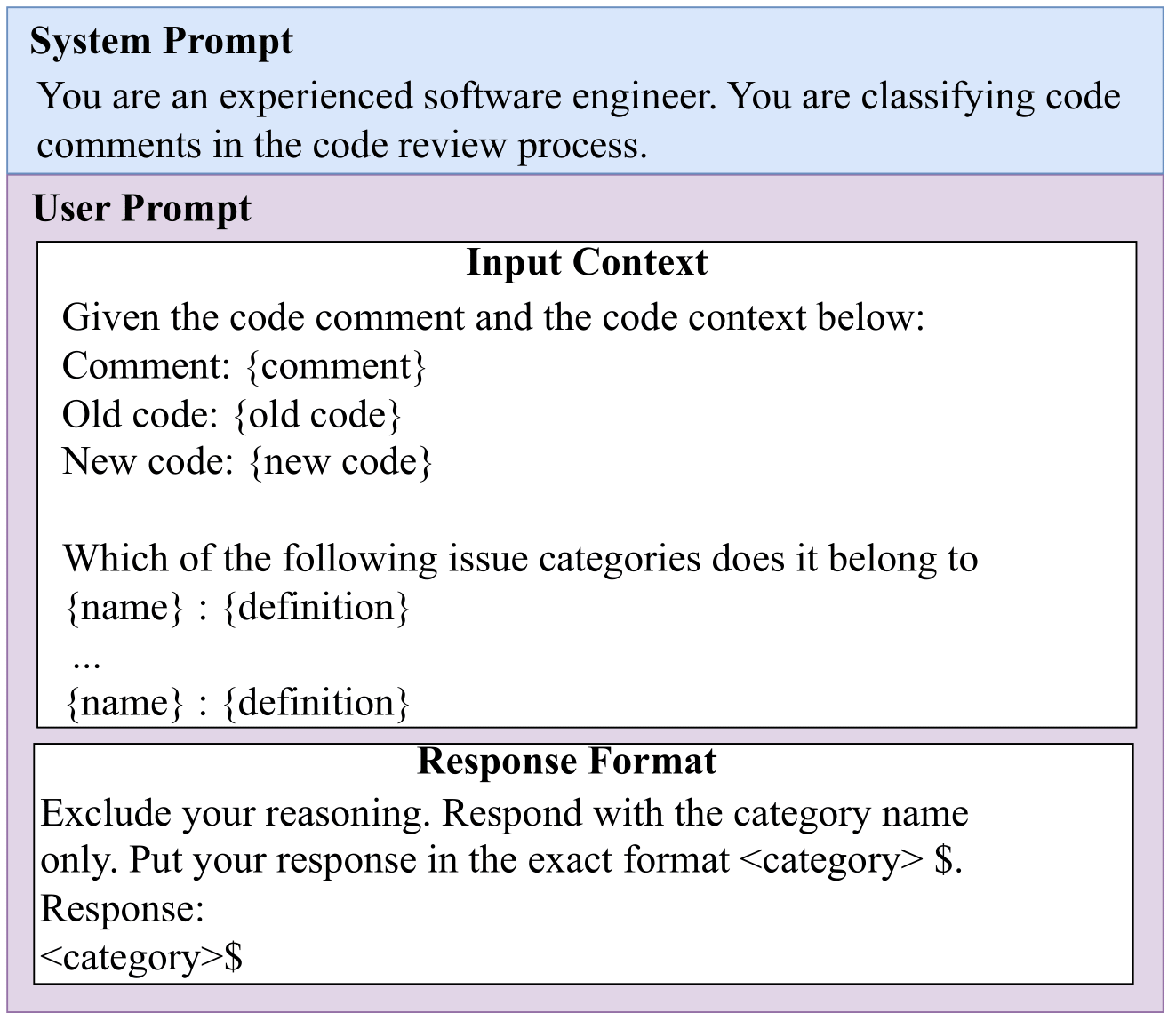}
    \caption{The prompt template that we used for classifying review comments with code as input context.}
    \label{fig:prompt_template}
\end{figure}

Figure \ref{fig:prompt_template} shows our prompt template for the review comment classification.
To instruct LLMs for effective classification, we design a prompt to be informative and concise, clearly outlining the task and providing sufficient context to guide the model’s decision-making~\cite{mu-etal-2024-navigating}.
The prompt instruction contains two main components, i.e., system and user prompts.
The system prompt specifies an overall objective of the task including role assignment.
The user prompt specifies a task instruction and answer format, which we describe in detail below.


\textbf{Task Instruction:} We formulate the classification task as a multiple choice question~\cite{robinson2023leveraging} to ensure that LLMs will select a category within the taxonomy.
Thus, we instruct LLMs to select a category based on the provided list.
The option list includes a pair of category (or group) names and their definitions, i.e., \texttt{\{name\}: \{definition\}}. 
To ensure that LLMs have sufficient information to understand each category, 
we refine the brief descriptions in the taxonomy of Turzo and Bosu~\cite{TurzoBosu} (see Table \ref{tab:review_comment_taxonomy}) with the detailed descriptions of Mäntylä and Lassenius~\cite{Mika}.\footnote{For categories present in Turzo and Bosu’s taxonomy~\cite{TurzoBosu_new} but absent in Mäntylä and Lassenius' work~\cite{Mika}, we retained the original descriptions unchanged.}
This refinement mainly involves adding elaborations and examples to improve clarity and better distinguish between categories.
For the definitions of the five high-level groups, we summarize the definitions of their child categories.
The complete definitions used in our prompt are provided in our replication package~\cite{our_replication_package}.

Context is important information that helps LLMs to determine the review comment category.
Thus, we explore two input context, i.e., with code change context (\textbf{code + comment}) and without code context (\textbf{comment only}).
Including the code context enables LLMs to analyze the relationship between the code and the review comment, enhancing their understanding of the potential concern expressed in the comment.
However, this also increases complexity and context length as LLMs must understand both code semantics and their relationship to the review comment.
Thus, we also evaluate the LLMs' performance without the code change context, i.e., removing \texttt{Old code: \{old code\}} and \texttt{New code: \{new code\}} out from the prompt template.
Note that \texttt{New code: \{new code\}} refers to the newly-authored code under review and \texttt{Old code: \{old code\}} refers to existing code in the repository.
When given only the review comment, LLMs will focus on  the linguistic features and comment intent to infer the classification.


\textbf{Prompt Response Format.}
\label{sec:response_standardization}
Since LLMs are generative models, they may generate responses that include additional details beyond the category name.   
Thus, our prompt includes an instruction for the response format (see Figure \ref{fig:prompt_template}). 
We use the \texttt{\$} symbol as a stopping criterion, allowing us to extract only the classification response without additional text.
It is possible that LLMs may still generate additional text before \texttt{\$} or even generate a classification response outside of the provided list. 
In such cases of any non-standard responses, we consider that none of the categories is selected, i.e., incorrect classification. 
Nonetheless, we observe that 
less than 5\% of instances in our experiments fall into these cases.




\section{Experimental Setup}
\label{sec:exp_setup}
\subsection{Dataset}

We use the dataset of the state-of-the-art approach~\cite{TurzoBosu_new}.
The dataset was originally obtained from the empirical study of Turzo and Bosu~\cite{TurzoBosu}, which contains 2,500 annotated code review comments from the OpenStack Nova Project, one of the largest open-source development communities with more than 15,000 contributors. 
The data was manually annotated by industry experts who reviewed both the associated code context and discussion threads and assigned labels using the taxonomy in Table~\ref{tab:review_comment_taxonomy}. 
Each category in the taxonomy is associated with a usefulness score ranked by OSS developers \cite{TurzoBosu}, making this dataset particularly suitable for evaluating LLMs' capability in classifying the usefulness of review categories.
As we want to explore the use of LLMs with code context, we exclude 672 comments in the dataset that do not have associated code changes.
Thus, we obtained 1,828 code review comments, which will be used throughout our experiments.
Table \ref{tab:review_comment_taxonomy} shows a distribution of comments across different categories.





\begin{table*}[htbp]
\centering
\caption{\label{tab:rq1_results_prompt_desgin}
The weighted average of classification results on 17 categories of our zero-shot LLM-based approaches.}
\begin{tabular}{ll|cccc|cccc}
        \toprule
         \textbf{Strategy} & \textbf{Model} & \multicolumn{4}{c}{\textbf{Comment Only}} & \multicolumn{4}{c}{\textbf{Code + Comment}} \\ \cmidrule(lr){3-6} \cmidrule(lr){7-10} 
        & & \textbf{F1} & \textbf{Precision} & \textbf{Recall} & \textbf{Accuracy} & \textbf{F1} & \textbf{Precision} & \textbf{Recall} & \textbf{Accuracy} \\ \midrule
 & Qwen 2-7B & 21.7 & 44.6 & 27.6 & 27.6 & 19.2 & 46.6 & 27.7 & 27.7 \\
 & Llama 3-8B & 20.2 & \textbf{47.3} & 26.1 & 26.1 & 21.9 & 52.1 & 28.5 & 28.5 \\
Flat &  Llama 3-70B & 33.5 & 42.1 & 35.9 & 35.9 & 38.7 & 49.4 & 40.2 & 40.2 \\
 & Qwen 2-72B & 31.5 & 44.7 & 29.5 & 29.5 & 34.9 & 46.8 & 36.3 & 36.3 \\
 & Llama 3.1-405B & \textbf{45.0} & 46.1 & \textbf{47.7} & \textbf{47.7} & \textbf{46.2} & 52.8 & \textbf{46.3} & \textbf{46.3} \\ 

\midrule \midrule
 & Qwen 2-7B & 31.9 & 39.1 & 34.6 & 34.6 & 29.2 & 38.8 & 34.0 & 34.0 \\
 & Llama 3-8B & 32.7 & 37.7 & 36.4 & 36.4 & 32.0 & 37.4 & 36.9 & 36.9 \\
Hierarchical &  Llama 3-70B& 36.8 & 40.7 & 39.1 & 39.1 & 40.1 & 43.9 & 41.1 & 41.1 \\
 & Qwen 2-72B & 38.3 & 41.4 & 38.9 & 38.9 & 40.1 & 44.9 & 41.6 & 41.6 \\
 &  Llama 3.1-405B& 40.3 & {45.4} & 41.4 & 41.4 & 42.8 & \textbf{53.6} & 43.6 & 43.6 \\ 
\midrule 
\multicolumn{2}{l|}{Baseline (Random)} & 6.8 & 11.7 & 5.4 & 5.4 & \multicolumn{4}{c}{-} \\

 \multicolumn{2}{l|}{Baseline (Majority class)} &   7.4 & 4.5& 21.1 & 21.2 & \multicolumn{4}{c}{-}  \\ 

\bottomrule

\end{tabular}
\end{table*}

\subsection{Studied LLMs}
To evaluate the capability of LLMs in classifying code review comments, we select recent models that have shown competency in both code and natural language-related tasks. 
Specifically, we examine two families of open-source instruction-tuned LLMs, i.e.,  Qwen 2~\cite{qwen2models} and Llama 3~\cite{llama3herdmodels}.
These models were trained on vast datasets including code and natural language.
Qwen 2 was trained on 7 trillion tokens and Llama 3 was trained on 15 trillion tokens.
Both models perform comparably to leading LLMs like GPT-4 across various tasks \cite{llama3herdmodels, qwen2models} and can be accessible via Hugging Face's Model Hub.

We explore LLMs with various sizes.
For Qwen 2 family, we use 7B and 72B.
For Llama 3 family, we use 8B, 70B, and 405B.  
The different model sizes allow us to balance performance and efficiency.
Large (405B) and medium models (70B/72B) excel at handling complex and longer context, while small models (8B/7B) offer faster inference times for resource-constrained environments.


We run the LLMs on our HPC machine, equipped with eight A100 GPUs and 500GB of RAM. However, as our machine cannot support the Llama 3.1-405B model, we access this model via TogetherAI’s API.\footnote{\url{https://www.together.ai/}}

\subsection{Evaluation Metrics}
\label{sec:evaluation_metrics}
To evaluate the classification effectiveness of LLMs, we use the metrics \textbf{F1-score}, \textbf{Precision}, \textbf{Recall}, and \textbf{Accuracy}. 
We calculate these metrics for each category individually where we treat the category of interest as positive and all other categories as negative.
To assess overall effectiveness across all categories, we calculate a \textbf{weighted average} given the imbalanced nature of our dataset.
The metric values for each category will be   proportional to its number of comments, which formulated as follows: 
\begin{align}
    \text{M}_{\text{weighted}} = \frac{\sum_{i}^{C} (\text{M}_i \times w_i)}{\sum_{i}^{C} w_i}
\end{align}
where $\text{M}_i \in \{precision, recall, F1, Accuracy\}$ for the category $i$ and $w_i$ is the proportion of comments belong to $i$.







\section{Experimental Results}    
\label{sec:results}
\subsection{(RQ1) Can we use LLMs to classify code review comments?  }

\textbf{Approach.} To answer RQ1, we conduct experiment with five LLMs (i.e., Qwen 2-7B, Llama 3-8B, Llama 3-70B, Qwen 2-72B, Llama 3.1-405B) using four prompts, i.e., \{flat, hierarchical\} strategies $\times$ \{comment only, code+comment\} context inputs.
As a result, we run a total of 20 experiments in classifying code review comments.
We run LLMs to classify all the 1,828 comments in the dataset into the 17 categories.

To estimate the lower bound performance of the review comment classification task, we use random guessing and majority.
These approaches do not require training data that are applicable for classifying review comments across the entire dataset like LLMs.
Note that this baseline only serves as a sanity check to determine if  LLMs can classify review comments. 
High-performing approaches will be further evaluated against the state-of-the-art approach in RQ2.

\textbf{Results.} Table~\ref{tab:rq1_results_prompt_desgin} presents our experimental results on classifying review comments using LLMs. 
LLMs achieve an average F1 score of 20.3\% - 46.2\% and an average Accuracy of 26.2\% - 47.7\%.
On the other hand, the baseline approaches only achieve an average F1 score of 6.8\% and 7.4\%; and an accuracy of 5.4\% and 21.2\%. 
This result indicates that LLMs have the potential in classifying review comments into 17 categories. 

Llama 3.1-405B consistently achieves high averaged F1 scores across different prompting approaches.
Specifically, Llama 3.1-405B with the flat strategy using code and comment achieves the highest average F1 score of 46.2\%.
Notably, Llama 3.1-405B performs relatively well even with review comments only using the \flatp{} strategy, achieving an average F1 score of 45\% and the highest average recall of 47.7\%. 


Providing code context with the review comment can improve the effectiveness for most of the models (7 out of 10), including Llama 3.1-405B. 
For example, for the medium-sized models (Llama 3-70B and Qwen 2-72B), a prompt with \textbf{code+comment} increase the F1-score by 8\% - 15\% (Llama 3-70B) to 4\% - 11\% (Qwen 2-72B) than a prompt with comment-only when using the \flatp{} and \hierarchical{} strategies, respectively.
On the other hand, the F1-scores of the small models (Llama 3-8B and Qwen 2-7B) are similar when using  prompts with and without code. The improvement in larger models when using additional context is partly due to their ability to handle longer contexts, which enables them to analyze the increased complexity of the code context and utilize additional information not present in the comment, resulting in better classification.

We observe that the small and medium models benefit from the hierarchical strategy.
Compared to the \flatp{} strategy, the small and medium models with the \hierarchical{} strategy consistently have better performance either using comment-only or code and comment in the prompt.
For example, when using the hierarchical strategy, the average F1-score of Qwen 2-7B is increased by 21\% (comment-only) and 14\% (code+comment), compared to the flat strategy.
On the other hand, Llama 3.1-405B has a performance drop when using the hierarchical strategy.
These results suggest that task decomposition is beneficial for small models, but may not be necessary for a large model as it can handle long input context and analyze holistic information on all categories.





\begin{tcolorbox}[size=title]
{\textbf{Answer to RQ1:}}
LLMs can be used to classify review comments into 17 categories, which is substantially better than the baseline approaches.
With the large model (Llama3.1-405B), using the flat strategy and providing code context achieves the highest average F1-score of 46.2.
For the small and medium models, using the hierarchical approach could help to boost their effectiveness.
\end{tcolorbox}

\subsection{(RQ2) Do LLMs outperform  the state-of-the-art approach? }

\begin{table*}[htbp]
\centering
\caption{Cross-validation results between LLMs and the state-of-the-art models.}
\setlength{\tabcolsep}{3pt}
\begin{tabular}{llcccc|cccc}
        \toprule
        \textbf{Model} &\multirow{2}{*}{\textbf{Strategy}}  & \multicolumn{4}{c}{\textbf{Comment Only}} & \multicolumn{4}{c}{\textbf{Code + Comment}} \\ \cmidrule(lr){3-6} \cmidrule(lr){7-10} 
        &  & \textbf{F1} & \textbf{Precision} & \textbf{Recall} & \textbf{Accuracy} & \textbf{F1} & \textbf{Precision} & \textbf{Recall} & \textbf{Accuracy} \\ \midrule
        
        Llama 3.1-405B &  Flat  & $\phantom{00}\textbf{45.0}^{**}$~~ & $\phantom{00}\textbf{46.8}^{***}$ & $\textbf{47.6}^{\circ}$ & $\textbf{47.6}^{\circ}$& 
        $\phantom{00}\textbf{46.7}^{**}$~~ & $\phantom{00}\textbf{51.9}^{**}$~~ & $46.6 ^{\circ}$& $46.6^{\circ}$  \\
        Llama 3.1-405B &  Hierarchical  & $40.1^{\circ}$  & $\phantom{0}45.0^{**}$ &	$41.4^{\circ}$  & $41.4^{\circ}$ & $44.8^{\circ}$ 	& $45.5^{\circ}$ 	& $44.9^{\circ}$  &  $45.6^{\circ}$\\ 
        Llama 3-70B &  Hierarchical  & $36.6^{\circ}$ & $41.3^{\circ}$ & $39.0^{\circ}$ & $39.0^{\circ}$ & $39.9^{\circ}$ & $44.0^{\circ}$ & $41.0^{\circ}$ & $41.0^{\circ}$ \\  
        Qwen 2-72B & Hierarchical  & $38.3^{\circ}$ & $42.0^{\circ}$ & $38.8^{\circ}$ & $38.8^{\circ}$ & $40.2^{\circ}$  & $44.6^{\circ}$ & $41.8^{\circ}$ & $41.8^{\circ}$ \\
         
        \midrule 
    \multicolumn{2}{l}{codeBERT+LSTM~\cite{TurzoBosu_new}} & 40.4~~ & 40.4~ & 45.5~ & 45.5~ & 42.4~~ & 43.8~ & 46.8~ & 46.8~ \\ 
        \bottomrule
    \multicolumn{10}{p{15cm}}{\footnotesize Statistical Significance: $p<0.001^{***}$, $p<0.01^{**}$, $p<0.05^{*}$, $p\geq0.05^{\circ}$ \cite{TurzoBosu_new} takes code + comment + features as input.} 
    \end{tabular}
    
\label{tab:model-comparison}
\end{table*}

\textbf{Approach.}
To answer RQ2, we compare the top-performing LLM-based approaches with the state-of-the-art approach~\cite{TurzoBosu_new}.
Based on the results in Table \ref{tab:rq1_results_prompt_desgin}, the four top-performing LLM-based approaches are Llama 3.1-405B with flat, Llama 3.1-405B with hierarchical, Qwen 2-72B with hierarchical, and Llama 3-70B with hierarchical.
Thus, for RQ2, we compare the performance of these four approaches with the state-of-the-art approach. 

To classify review comments, Turzo et al.~\cite{TurzoBosu_new} uses CodeBERT–a
pre-trained language model to generate a representation of a
code review comment then train an LSTM model to classify
comments.
This approach has shown superior results compared to traditional machine learning-based methods~\cite{Fregnan-etal-2022-classifying} e.g., support vector machines in categorizing review comments into the five high-level groups in the defect-based taxonomy.
As our work is the first to classify 17 categories, the CodeBERT+LSTM approach by Turzo et al.~\cite{TurzoBosu_new} is closely relevant to our work, and thus serves as our state-of-the-art baseline.
We use the replication package provided by Turzo et al. to classify review comments by adjusting only the five high-level groups to 17 categories.
We use their best-performing approach, i.e., using code, comment, and features as input.
We use the same hyperparameter settings as the prior work.

Since the CodeBERT+LSTM approach requires training, we conducted an experiment using 10-fold cross-validation, following the same experimental setup as the prior work~\cite{TurzoBosu_new}.
In particular, we use nine folds for training and the remaining fold for evaluation.
We repeat this process for every fold and measure an average performance. 
As LLMs do not require training, we directly conduct inference by using the same test fold and measure the average results.
To measure statistical significance, we use the one-sided Wilcoxon signed-rank test.
We compare the performances across all folds between each LLM-based approach and the CodeBERT+LSTM baseline.

\textbf{Results.}
Table~\ref{tab:model-comparison} shows the results 
based on cross-validation.
We find that Llama 3.1-405B outperforms the state-of-the-art approach.
On average, the CodeBERT+LSTM approach achieves an F1 score of 40.4 (comment-only) and 42.4 (code+comment).
Llama 3.1-405B with the flat strategy achieves an F1 score of 45 (comment-only) and 46.7 (code+comment), achieving an increase of 11.3\% and 10\%.
The statistical test also confirms that the F1 score Llama 3.1-405B is significantly higher than the CodeBERT+LSTM approach.
On the other hand, Llama 3.1-405B with the hierarchy strategy also achieves a comparable result as the CodeBERT+LSTM approach.

The performance of the medium models do not surpass the state-of-the-art approach.
Table \ref{tab:model-comparison} shows that Llama 3-70B and Qwen 2-72B with the hierarchical strategy achieve lower F1 scores than the baseline approach with a small difference.
For example, Llama 3-70B and  Qwen 2-72B using code+comment achieve F1 scores 5.8\% and 5\% lower than the baseline approach, respectively.

\begin{tcolorbox}[size=title]
{\textbf{Answer to RQ2:}}
LLMs can outperform the state-of-the-art model, i.e., a supervised deep learning approach. 
When using Llama 3.1-405B with the flat strategy, the F1 score is significantly increased by 11.3\% and 10\% when without and with code, respectively.
Nonetheless, using the smaller models will result in a lower performance than the supervised approach.
\end{tcolorbox}


\subsection{(RQ3) Which categories can LLMs accurately classify?}
\begin{table*}[!t]
\centering
\setlength{\tabcolsep}{4pt}
\caption{\label{table:category_comparison}Category-wise classification performance.}
\resizebox{\textwidth}{!}{
\begin{tabular}{l|r|r|r|r|r|r|p{0.6cm}|p{0.6cm}|p{0.6cm}|c}
\toprule
\multicolumn{1}{c|}{\textbf{Category}} & \multicolumn{3}{c|}{\textbf{Llama 3-405B (Flat, Code+Comment)}} & \multicolumn{3}{c|}{\textbf{Qwen 2-72B (Hierarchical, Code+Comment)}} & \multicolumn{3}{c|}{\textbf{CodeBERT + LSTM}} & \\\cmidrule{2-10}
 & \textbf{F1} & \textbf{Prec} & \textbf{Rec} & \textbf{F1} & \textbf{Prec} & \textbf{Rec} & \textbf{F1} & \textbf{Prec} & \textbf{Rec} & \textbf{\#Comments} \\ \midrule 
Functional Defect (Most Useful)$^\dag$ & \textbf{10.4} (n/a) & \textbf{6.8} (n/a)& \textbf{25.0} (n/a) & \textbf{7.9} (n/a) & \textbf{4.6} (n/a) & \textbf{30.0} (n/a) & 0.0 & 0.0 & 0.0 & 12 \\
Validation & \textbf{21.8} (29\%) & \textbf{44.7} (124\%) & 15.0 (-18\%) & \textbf{24.9} (47\%) & \textbf{36.5} (83\%) & \textbf{22.5} (22\%)& 16.9 & 19.9 & 18.5 & 90 \\
Logical & \textbf{13.2} (n/a) & \textbf{16.7} (n/a) & \textbf{12.7} (n/a) & \textbf{10.0} (n/a) & \textbf{10.1} (n/a) & \textbf{10.1} (n/a) & 0.0 & 0.0 & 0.0 & 56 \\
Interface & 0.0 & 0.0 & 0.0 & \textbf{12.7} (535\%) & \textbf{10.3} (312\%) & \textbf{20.0} (1075\%) & 2.0 & 2.5 & 1.7 & 30 \\
Solution Approach & \textbf{47.9} (28\%) & \textbf{45.6} (26\%) & \textbf{52.2} (20\%) & \textbf{42.0} (12\%) & 34.1 (-5\%) & \textbf{56.0} (28\%) & 37.4 & 36.0 & 43.6 & 201 \\
Question & 66.1 (-.3\%) & \textbf{64.1} (2\%) & 68.5 (-8\%) & 42.5 (-36\%) & 55.0 (-12\%) & 34.8 (-53\%)  & 66.3 & 62.8 & 74.1 & 275 \\
Design Discussion & \textbf{26.9} (88\%) & \textbf{29.8} (58\%) & \textbf{26.1} (114\%) & 13.9 (-2.8\%) & 16.0 (-15\%) & \textbf{13.3} (9\%) & 14.3 & 18.8 & 12.2 & 87 \\
Resource & 0.0 & 0.0 & 0.0 & \textbf{2.0} (n/a) & \textbf{5.0} (n/a) & \textbf{1.2} (n/a) & 0.0 & 0.0 & 0.0 & 34 \\
Documentation & 52.8 (-18\%) & 44.1 (-20\%) & 66.4 (-16\%) &63.7 (-2\%) & \textbf{57.5} (3\%) & 72.4 (-9\%) & 65.0 & 55.6 & 79.6 & 387 \\
Organization of Code & \phantom{.}\textbf{50} (51\%) & \textbf{45.4} (44\%) & \textbf{57.2} (32\%) & \textbf{49.0} (48\%) & \textbf{42.3} (34\%) & \textbf{60.2} (39\%)& 33.1 & 31.6 & 43.3 & 184 \\
Alternate Output & \textbf{40.8} (213\%) & \textbf{54.7} (219\%) & \textbf{34.6} (212\%) &10.0 (-23\%) & \textbf{25} (46\%) & 6.5 (-41\%) & 13.0 & 17.1 & 11.1 & 64 \\
Support & \textbf{10.0} (n/a) & \textbf{15.0} (n/a) & \textbf{12.0} (n/a) &0.0 & 0.0 & 0.0 & 0.0 & 0.0 & 0.0 & 14 \\
Timing & \textbf{15.0} (n/a) & \textbf{13.3} (n/a) & \textbf{20.0} (n/a) & \textbf{16.7} (n/a) & \textbf{20.0} (n/a) & \textbf{15.0} (n/a) & 0.0 & 0.0 & 0.0 & \phantom{00}4 \\
Naming Convention & \textbf{57.8} (107\%) & \textbf{76.2} (55\%) & \textbf{47.3} (80\%) & \textbf{54.0} (94\%) & \textbf{72.1} (47\%) & \textbf{44.9} (71\%) & 27.9 & 49.0 & 26.3 & 76 \\
Praise & \textbf{62.6} (5\%) & 65.0 (-1\%)  & \textbf{62.8} (6\%) & 40.3 (-33\%) & \textbf{66.7} (.8\%) & 33.4 (-44\%) & 59.8 & 66.2 & 59.2 & 83 \\
Visual Representation (Least Useful)$^\dag$ & \textbf{71.4} (107\%) & \textbf{80.4} (40\%) & \textbf{65.6} (149\%) & \textbf{70.7} (105\%) & \textbf{86.8} (52\%) & \textbf{62.3} (137\%) & 34.4 & 57.3 & 26.3 & 73 \\ \bottomrule 
\multicolumn{8}{p{13cm}}{\footnotesize \textbf{Bold} text indicates higher performance than the state-of-the-art approach. }\\

\multicolumn{8}{p{13cm}}{\footnotesize $^\dag$ The categories are sorted in terms of usefulness ratings, as rated by practitioners~\cite{TurzoBosu} (see Table \ref{tab:review_comment_taxonomy}). }
\end{tabular}
}
\end{table*}

\textbf{Approach} To address RQ3, we evaluate the performance of  LLMs and the state-of-the-art approach for each category.
Instead of using a weighted average, we examine the F1-score, Precision, Recall, and Accuracy for each category individually. 
To compare with the state-of-the-art approach, we analyze the cross-validation classification results.
We compare the category-wise performance of Llama 3.1-405B with the flat strategy and Qwen 2-72B with the hierarchical strategy, using code + comment, against the state-of-the-art approach (CodeBERT+LSTM)~\cite{TurzoBosu_new}.
Llama 3.1-405B with the flat strategy using code + comment is the best-performing LLM-based approach.
Although the medium-sized models do not outperform the state-of-the-art approach overall, exploring their potential benefits in specific categories could be valuable.

To compare the performance, we quantify the percentage difference using the calculation of $\frac{M_\mathrm{our} - M_{baseline}}{M_{basline}}\times 100\%$, where the averaged results of $\text{M}_i \in \{precision, recall, F1\}$.

\textbf{Results} Table~\ref{table:category_comparison} shows the category-wise classification performance between our LLM-based approaches and the state-of-the-art approach.
The bold text indicates that the performance of our LLM-based approach is higher than the CodeBERT+LSTM approach.
The categories are shown in descending order based on the usefulness ranking. 

LLMs are effective at classifying praise and visual representation with an F1 score of 62.6 and 71.4, respectively.
Moreover, LLMs outperform the state-of-the-art approach in the five most useful categories (i.e., functional defect, validation, logical, interface, and solution approach). 
Table~\ref{table:category_comparison} shows Llama 3.1-405B and Qwen 2-72B outperform for four and five most useful categories, respectively.
Notably, in the functional defect and logical categories, which the prior approach fails to identify, our Llama 3.1-405B achieves F1 scores of 10.4\% and 13.2\%, respectively. 
For the validation and solution approach categories, our LLM-based approach also achieves an F1 score of 47\% (Qwen 2-72B) and 28\% (Llama 3.1-405B) higher than the state-of-the-art approach.
Our LLM-based approach with Qwen 2-72B also can improve the F1 score for the interface category by 535\%. 
Overall, our LLM-based approach can generally achieve higher F1 scores for 10 (Qwen2-7B) and 12 (Qwen 2-72B) out of 17 categories.
These results suggest that our LLM-based approaches are more effective at identifying useful code review comments.

LLMs can classify review comment categories where the state-of-the-art approach struggles, i.e., functional defect, logical, support, and timing.
These are challenging categories as they have a low proportion of comments, but we observe a substantial improvement for these low-resource categories when using LLM-based approaches.
For example, the design discussion, alternate output, and naming convention categories have only 87, 64, and 76 comments in the dataset.
However, our LLM-based approach (LLama 3.1-405B) achieves F1 scores 88\%, 213\%, and 107\% higher than the state-of-the-art approach.
These results show that our LLM-based approach is effective in handling categories with a low proportion of comments, whereas traditional supervised methods may struggle due to insufficient data.


\begin{tcolorbox}[size=title]
{\textbf{Answer to RQ3:}}
When examining performance across each category,  LLMs substantially outperform the state-of-the-art method in the five most important categories, i.e., functional defect, logical, validation, interface, and solution approach. Additionally, LLMs have shown a considerable improvement in categories where the state-of-the-art method struggles due to insufficient data.
\end{tcolorbox}


\section{Discussion \& Suggestions}
\label{sec:discussion}
In this section, we discuss the performance of  LLMs in various aspects and provide suggestions. 

\subsection{Performance of LLMs}
As the prompt used in this study contains two main components (i.e., input information and classification strategies), we now further examine the potential impact of these components on the performance of the models.

\textbf{The category definition.}
\textit{Is the performance of LLMs impacted when using the original brief definitions?}
As described in Section \ref{sec:llm_approach}, we elaborated the category definitions from Turzo and Bosu \cite{TurzoBosu} based on the taxonomy of Mäntylä and Lassenius \cite{Mika} to provide sufficient information to the LLMs. 
However, as our results show (flat vs hierarchical), long input contexts can affect the LLM performance. The length of the category definitions may also have an impact; therefore, we examine whether it impacts the model's effectiveness.

To measure this impact, we use the original brief description of categories from Turzo and Bosu \cite{TurzoBosu} instead of our refined definitions.
Then, we instruct LLMs with code+comment to classify the review.
We then measure the percentage difference when changing from our refined definitions to the brief definitions using a calculation of  $\frac{M_o - M_r}{M_r} \times 100\%$, where $M_o$ is the performance when using the original brief definitions and $M_r$ is the performance when using our refined definitions.
Table~\ref{tab:model-performance-delta} shows that changing to the original brief definitions has a negligible to small impact on large and medium-sized models (0.2\% - 7.5\% changes in F1 scores). 
On the other hand, changing a definition has a considerable impact on smaller-sized models.
Interestingly, while the original brief definitions negatively impact Qwen 2-7B, they improve the F1 score of Llama 3-8B by 20.2\% (in the hierarchical strategy).
These results suggest that, when using smaller models for classification tasks, carefully optimizing the information provided is essential to maximize performance.

\begin{table}[!t]
\centering
\setlength{\tabcolsep}{4.5pt}
\caption{\label{tab:model-performance-delta}The percentage change when using the original brief definitions of Turzo and Bosu~\cite{TurzoBosu} instead of using our refined definitions.
The results are based on code + comment as input.}
\begin{tabular}{llcccc}
    \toprule
    \textbf{Strategy} & \textbf{Model} & \textbf{F1} & \textbf{Precision} & \textbf{Recall} & \textbf{Accuracy} \\
    \midrule
    \parbox[t]{2mm}{\multirow{5}{*}{\rotatebox[origin=c]{90}{\textbf{Flat}}}} & Qwen 2-7B & -33.0\% & -2.8\% & -12.3\% & -12.3\% \\
                          & Llama 3-8B & 10.9\% & -23.2\% & 5.4\% & 5.4\% \\
                          & Llama 3-70B & 7.5\% & -5.1\% & 4.2\% & 4.2\% \\
                          & Qwen 2-72B & 6.3\% & 0.9\% & -0.5\% & -0.5\% \\
                          & Llama 3.1-405B & -0.9\% & -9.7\% & -1.1\% & -1.1\% \\ \midrule 
                          & Average & -1.9\% & -8.0\% & -0.9\% & -0.9\% \\
    \midrule
    \parbox[t]{2mm}{\multirow{5}{*}{\rotatebox[origin=c]{90}{\textbf{Hierarchical}}}} & Qwen 2-7B & -10.3\% & 6.4\% & -5.9\% & -5.9\% \\
                                  & Llama 3-8B & 20.2\% & 2.8\% & 17.1\% & 17.1\% \\
                                  & Llama 3-70B & -0.5\% & 3.0\% & -1.9\% & -1.9\% \\
                                  & Qwen 2-72B & -0.2\% & 1.6\% & -1.0\% & -1.0\% \\
                                  & Llama 3.1-405B & -3.0\% & -4.8\% & -3.4\% & -3.4\% \\ \midrule 
                                  & Average & 1.2\% & 1.8\% & 1.0\% & 1.0\% \\
    \bottomrule

\end{tabular}

\end{table}

\textbf{The \hierarchical{} strategy.}
\textit{What is the accuracy of the high-level classification in the hierarchical strategy?}
In the hierarchical strategy, we employ a two-step process: 1) We classify review comments into one of the five high-level groups, then 2) we classify them into the specific categories within the predicted group.
Our RQ1 results show that the \hierarchical{} strategy can improve model performance for small and medium-sized models compared to the \flatp{} strategy. This is partly due to the reduced complexity of decomposing the task into two steps. 
However, there is a risk that incorrect classification from the step 1 (i.e., a high-level group classification) will propagate into the second step. 
To gain further insights, we analyze the performance of step 1 in our hierarchical strategy, i.e., classifying five high-level groups.
Note that we only examine the models with code+comment as input. 

We found that Llama 3.1-405B achieves an F1 score of 51.9 and an accuracy of 54.8. 
While the small models perform lower, with F1 scores of 44.0 (Llama 3-8B) and 39.6 (Qwen 2-7B), the medium-sized models show results comparable to the large model, achieving F1 scores of 50.1 (Qwen 2-72B) and 48.8 (Llama 3-70B). 
These findings suggest that the performance of medium-sized models could be further improved by optimizing the high-level classification, potentially providing a good balance between computational efficiency and accuracy—making them a practical choice for resource-constrained environments.

\textbf{Inference Time.} The runtime for LLama3-405B, medium models (72B), and small models (7-8B) was 9, 5, and 1 hours, respectively, while the CodeBERT+LSTM approach took 5 hours.
Although LLMs have longer runtimes, they offer a scalable solution by eliminating time-consuming manual annotation and training time. 
As shown in Table ~\ref{table:category_comparison}, medium-sized models like Qwen 2-72B offer an optimal balance between runtime and performance, achieving relatively high F1 scores on imbalanced datasets with the same runtime as the CodeBERT+LSTM approach.





\subsection{Advantage and Practical Usage of LLMs for fine-grained comment classification}
We now discuss the advantages of our LLM-based approaches in classifying 17 categories and outline the potential usage of our approach for both practitioners and researchers.

\textbf{Advantage.}
Our findings from RQ2 show that LLMs can outperform the state-of-the-art, which is a trained deep learning model. Additionally, RQ3 demonstrates that  LLMs are particularly effective in handling categories with a low proportion of comments, where the state-of-the-art approach struggles due to limited data.
These results highlight the advantages of leveraging LLMs’ pre-trained knowledge through prompt engineering, which mitigates challenges associated with limited labeled data in code review classification—an issue that can impact such a supervised deep learning model like CodeBERT+LSTM. While CodeBERT+LSTM achieved high accuracy and recall overall, it showed significant bias toward majority classes, scoring near-zero on minority classes such as timing, functional defect, support, and interface.
In contrast, LLMs demonstrated more balanced performance across both high- and low-frequency categories, counteracting class imbalance issues, making these LLMs particularly effective for datasets with uneven category distributions. 


\textbf{Practical Usage Scenarios.} 
Based on the study results, the ability of our LLM-based approaches could offer a scalable solution for code review analytics and code review automation.

\textit{Code review analytics}: The LLM-based approach can improve the effectiveness of the CR process by analyzing the nature of review comments. 
The task of addressing code review comments often induces heavy work loads and require context switching~\cite{Czerwonka,Oleksii2016}.
As such, it is imperative to ease this load by helping developers focus on the most critical issues first, rather than subjecting them to all code reviews at once.
Based on our RQ3 findings, our LLM-based approach is effective at classifying praise and visual representation comments which are ranked as the least useful by practitioners. 
The current LLM capabilities can help teams assess and identify these less useful comments. 
For example, an LLM classifier can be integrated into the code review platform to tag and hide less useful comments as they are being left by the reviewers, surfacing only more important code reviews. 
When inspecting the comments left on the platform, the busy developers will then be faced with a reduced set of only the most urgent issues.
The developers can then choose to surface the remaining trivial comments at a later time when they have more capacity.

\textit{Code review automation}: Several studies have developed approaches for code review automation such as automated review comment generation (i.e., generating a comment given a code change) ~\cite{Li2022CodeReviewer,Tufano2022PretrainedModels,Lin2024} and automated code refinement (i.e., generating a code revision given a code change and comment) ~\cite{tufano2019learning,thongtanunam2022autotransform,pornprasit2023d}.
These approaches train the neural models using large-scale code review datasets mined from open-source platforms like Gerrit and GitHub. 
However, they typically treat all comment types as equally valuable. 
Given that practitioners perceive certain comment types as more useful than others~\cite{TurzoBosu}, our LLM-based approach sheds light on a new direction for code review automation—allowing for a focus on more useful comment types. 
For example, practitioners may prefer to have generated comments related to technical concerns, rather than visual representation which can be detected by linters~\cite{Oleksii2016,hasan}.
Code review automation research can leverage our LLM-based approach to remove comments considered to be less useful from the training and evaluation data, thus improving both the practicality and quality of their approaches.



\subsection{Suggestions for Future Work}
We now discuss the performance of LLMs in the challenging categories and potential improvement for future work.

\textbf{Challenging Categories.}
RQ3 shows that LLMs underperforms for  \textit{Support}, \textit{Resource}, and \textit{Interface} categories, indicating that these categories are challenging for our approach.
This is partly due to various factors.
One possible reason is due to the common characteristic of the review comments which are short and context-dependent, especially for the interface category which is related to the interaction with other parts of the system (e.g., APIs).
For example, a comment ``\textit{We should have already set it in the API before we hit this.}'' requires an understanding of its code repository to infer the issue type.
Another possible reason is the nuanced nature of the comment, which can be difficult to interpret, for example, a comment ``\textit{Nit: prefer parens over backslash}''.
This suggests a need for techniques that enhance clarity and provide more contextual information to help LLMs better understand review comments, leading to improved classification performance.

The definitions of the categories may also play a significant role. 
For example, the Support, Resource, and Interface categories address similar parts of the system (e.g., libraries) but from different perspectives. 
For instance, Support focuses on the configuration of libraries or other components, while Interface emphasizes the interactions between them. 
Thus, low LLM performance may also stem from comments that can plausibly fit under multiple similar categories when not considering subtle distinctions.
Therefore, future work should refine the category definitions, ensuring they highlight distinct characteristics for better clarity and differentiation.

\textbf{More advanced techniques for code review classification.}
This work has explored prompting approaches with varied input information.
Future work can explore advanced prompting techniques to better leverage in-context learning. This includes designing chain-of-thought prompts \cite{Jason-etal-Chain}, which break down classification tasks into intermediate steps, and experimenting with few-shot or retrieval augmented prompting \cite{Brown-etal-2020-few-shot}, which provides examples to help LLMs better understand class definitions. 
Unfortunately, with a limited and imbalanced benchmark dataset, there are not sufficient examples to conduct experiment with these techniques. 
Moreover, these prompting approaches require a careful design, as our study results show that long input length potentially has a negative impact on model performance.

Additionally, data scarcity and imbalance issues have been hindering progress in this direction. Future work can explore using knowledge distillation from LLMs to automatically label more data \cite{pavlovic-poesio-2024-effectiveness} and increase the dataset size that can be used to fine-tune smaller language models \cite{pangakis-wolken-2024-knowledge} that can achieve higher performance while providing more efficiency.








\section{Threats to Validity}
\label{sec:threat}

This section addresses potential threats to the validity.

\textbf{Internal Validity.} 
To make a fair comparison with the state-of-the-art approach\cite{TurzoBosu_new}, we adapted their publicly available code with minimal modifications to classify comments into 17 level-2 categories instead of the original 5 level-1 categories. To accomplish this, we adjusted the model’s output layer to predict among the 17 subcategories and modified the input dictionary to include or exclude code context as needed. Furthermore, we preserved their 10-fold cross-validation approach, minimizing risks related to the replication process.

\textbf{Construct Validity.} 
Due to the inherent variability in LLM outputs, particularly in smaller models, responses may sometimes deviate from prompt instructions even when models are instruction-tuned and the prompting hyperparameters are fixed. 
In our experiments, we used the special character ``\$" to prompt the LLMs to stop generating tokens after providing a classification label. However, occasionally, small models generate extraneous text or out-of-list labels. In these instances, we assigned a default label of ``False Positive'' to all categories, as outlined in Section~\ref{sec:response_standardization}. This affected fewer than 5\% of instances, which is relatively minor.. 
Other approaches for generating classification labels may yield different results.

\textbf{External Validity.} 
A threat to external validity arises from the use of manually labeled data from Turzo et al.'s studies \cite{TurzoBosu, TurzoBosu_new}. Although labeled by experts, manual interpretation may vary, introducing potential biases or inconsistencies. 
The annotation agreement was reported as 0.68 (Cohen’s $\kappa$), indicating some annotation ambiguity exist, potentially impacting the generalizability of our findings.
We addressed this by adhering closely to the taxonomy provided and ensuring consistency in label interpretation. However, the alignment of categories with broader CR contexts or alternative datasets remains a potential limitation.

\section{Conclusion}
\label{sec:conclusion}
The quality of code review (CR) largely depends on reviewer comments, which are ideally constructive and actionable. 
Automated CR comment classification not only offers insights into CR practices, but also enhances CR effectiveness. 
While prior research has explored automated classification of CR comments, none have examined methods that capture the underlying concerns at a more fine-grained level.

In this work, we explore the capability of LLMs to classify code review comments into 17 categories.
Our results show that the best-performing LLMs can classify review comments with an average F1 score of  46.2\%.
Moreover, we found that LLMs can substantially outperform the state-of-the-art method in the five most useful categories, where the state-of-the-art method struggles due to insufficient data.
Our work highlights the advantages of leveraging LLMs. 
Specifically, our LLM-based approach demonstrated more balanced performance across both high and low frequency categories, counteracting the class imbalance issues.
LLMs could offer a scalable solution for code review analytics to improve the
effectiveness of the CR process.

\bibliographystyle{IEEEtran}
\bibliography{egbib}

\begin{thebibliography}{10}
\providecommand{\url}[1]{#1}
\csname url@samestyle\endcsname
\providecommand{\newblock}{\relax}
\providecommand{\bibinfo}[2]{#2}
\providecommand{\BIBentrySTDinterwordspacing}{\spaceskip=0pt\relax}
\providecommand{\BIBentryALTinterwordstretchfactor}{4}
\providecommand{\BIBentryALTinterwordspacing}{\spaceskip=\fontdimen2\font plus
\BIBentryALTinterwordstretchfactor\fontdimen3\font minus \fontdimen4\font\relax}
\providecommand{\BIBforeignlanguage}[2]{{%
\expandafter\ifx\csname l@#1\endcsname\relax
\typeout{** WARNING: IEEEtran.bst: No hyphenation pattern has been}%
\typeout{** loaded for the language `#1'. Using the pattern for}%
\typeout{** the default language instead.}%
\else
\language=\csname l@#1\endcsname
\fi
#2}}
\providecommand{\BIBdecl}{\relax}
\BIBdecl

\bibitem{Bacchelli}
A.~Bacchelli and C.~Bird, ``Expectations, outcomes, and challenges of modern code review,'' in \emph{2013 35th International Conference on Software Engineering (ICSE)}, 2013, pp. 712--721.

\bibitem{Bosu2016}
A.~Bosu, J.~C. Carver, C.~Bird, J.~Orbeck, and C.~Chockley, ``Process aspects and social dynamics of contemporary code review: Insights from open source development and industrial practice at microsoft,'' \emph{IEEE Transactions on Software Engineering}, vol.~43, no.~1, pp. 56--75, 2017.

\bibitem{Bird}
C.~Bird, T.~Carnahan, and M.~Greiler, ``Lessons learned from building and deploying a code review analytics platform,'' in \emph{2015 IEEE/ACM 12th Working Conference on Mining Software Repositories}, 2015, pp. 191--201.

\bibitem{Czerwonka}
J.~Czerwonka, M.~Greiler, and J.~Tilford, ``Code reviews do not find bugs. how the current code review best practice slows us down,'' in \emph{2015 IEEE/ACM 37th IEEE International Conference on Software Engineering}, vol.~2, 2015, pp. 27--28.

\bibitem{thongtanunam2015investigating}
P.~Thongtanunam, S.~McIntosh, A.~E. Hassan, and H.~Iida, ``Investigating code review practices in defective files: An empirical study of the qt system,'' in \emph{Proceedings of the International Conference on Mining Software Repositories}.\hskip 1em plus 0.5em minus 0.4em\relax IEEE, 2015, pp. 168--179, acceptance Rate: 30

\bibitem{Bosu2015}
A.~Bosu, M.~Greiler, and C.~Bird, ``Characteristics of useful code reviews: an empirical study at microsoft,'' in \emph{Proceedings of the 12th Working Conference on Mining Software Repositories}, ser. MSR '15.\hskip 1em plus 0.5em minus 0.4em\relax IEEE Press, 2015, p. 146–156.

\bibitem{hasan}
\BIBentryALTinterwordspacing
M.~Hasan, A.~Iqbal, M.~R.~U. Islam, A.~J. M.~I. Rahman, and A.~Bosu, ``Using a balanced scorecard to identify opportunities to improve code review effectiveness: An industrial experience report,'' \emph{CoRR}, vol. abs/2101.10585, 2021. [Online]. Available: \url{https://arxiv.org/abs/2101.10585}
\BIBentrySTDinterwordspacing

\bibitem{Rahman}
\BIBentryALTinterwordspacing
M.~M. Rahman, C.~K. Roy, and R.~G. Kula, ``Predicting usefulness of code review comments using textual features and developer experience,'' in \emph{Proceedings of the 14th International Conference on Mining Software Repositories}, ser. MSR '17.\hskip 1em plus 0.5em minus 0.4em\relax IEEE Press, 2017, p. 215–226. [Online]. Available: \url{https://doi.org/10.1109/MSR.2017.17}
\BIBentrySTDinterwordspacing

\bibitem{TurzoBosu}
\BIBentryALTinterwordspacing
A.~K. Turzo and A.~Bosu, ``What makes a code review useful to opendev developers? an empirical investigation,'' \emph{Empirical Softw. Engg.}, vol.~29, no.~1, Nov. 2023. [Online]. Available: \url{https://doi.org/10.1007/s10664-023-10411-x}
\BIBentrySTDinterwordspacing

\bibitem{pangsakulyanont2014assessing}
T.~Pangsakulyanont, P.~Thongtanunam, D.~Port, and H.~Iida, ``Assessing mcr discussion usefulness using semantic similarity,'' in \emph{Proceedings of the International Workshop on Empirical Software Engineering in Practice}, 2014, pp. 49--54.

\bibitem{ebert2017confusion}
F.~Ebert, F.~Castor, N.~Novielli, and A.~Serebrenik, ``Confusion detection in code reviews,'' in \emph{2017 IEEE International Conference on Software Maintenance and Evolution (ICSME)}.\hskip 1em plus 0.5em minus 0.4em\relax IEEE, 2017, pp. 549--553.

\bibitem{Yang-etal-2023-EvaCRC}
\BIBentryALTinterwordspacing
L.~Yang, J.~Xu, Y.~Zhang, H.~Zhang, and A.~Bacchelli, ``Evacrc: Evaluating code review comments,'' in \emph{Proceedings of the 31st ACM Joint European Software Engineering Conference and Symposium on the Foundations of Software Engineering}, ser. ESEC/FSE 2023.\hskip 1em plus 0.5em minus 0.4em\relax New York, NY, USA: Association for Computing Machinery, 2023, p. 275–287. [Online]. Available: \url{https://doi.org/10.1145/3611643.3616245}
\BIBentrySTDinterwordspacing

\bibitem{Fregnan-etal-2022-classifying}
\BIBentryALTinterwordspacing
E.~Fregnan, F.~Petrulio, L.~Di~Geronimo, and A.~Bacchelli, ``What happens in my code reviews? an investigation on automatically classifying review changes,'' \emph{Empirical Softw. Engg.}, vol.~27, no.~4, 2022. [Online]. Available: \url{https://doi.org/10.1007/s10664-021-10075-5}
\BIBentrySTDinterwordspacing

\bibitem{TurzoBosu_new}
\BIBentryALTinterwordspacing
A.~K. Turzo, F.~Faysal, O.~Poddar, J.~Sarker, A.~Iqbal, and A.~Bosu, ``{ Towards Automated Classification of Code Review Feedback to Support Analytics },'' in \emph{2023 ACM/IEEE International Symposium on Empirical Software Engineering and Measurement (ESEM)}.\hskip 1em plus 0.5em minus 0.4em\relax Los Alamitos, CA, USA: IEEE Computer Society, Oct. 2023, pp. 1--12. [Online]. Available: \url{https://doi.ieeecomputersociety.org/10.1109/ESEM56168.2023.10304851}
\BIBentrySTDinterwordspacing

\bibitem{Mika}
M.~V. Mäntylä and C.~Lassenius, ``What types of defects are really discovered in code reviews?'' \emph{IEEE Transactions on Software Engineering}, vol.~35, no.~3, pp. 430--448, 2009.

\bibitem{Chang-etal-2024-surveyLLM}
\BIBentryALTinterwordspacing
Y.~Chang, X.~Wang, J.~Wang, Y.~Wu, L.~Yang, K.~Zhu, H.~Chen, X.~Yi, C.~Wang, Y.~Wang, W.~Ye, Y.~Zhang, Y.~Chang, P.~S. Yu, Q.~Yang, and X.~Xie, ``A survey on evaluation of large language models,'' \emph{{ACM} Trans. Intell. Syst. Technol.}, vol.~15, no.~3, pp. 39:1--39:45, 2024. [Online]. Available: \url{https://doi.org/10.1145/3641289}
\BIBentrySTDinterwordspacing

\bibitem{Zhao-etal-2023-LLMSurvey}
\BIBentryALTinterwordspacing
W.~X. Zhao, K.~Zhou, J.~Li, T.~Tang, X.~Wang, Y.~Hou, Y.~Min, B.~Zhang, J.~Zhang, Z.~Dong, Y.~Du, C.~Yang, Y.~Chen, Z.~Chen, J.~Jiang, R.~Ren, Y.~Li, X.~Tang, Z.~Liu, P.~Liu, J.-Y. Nie, and J.-R. Wen, ``A survey of large language models,'' \emph{arXiv preprint arXiv:2303.18223}, 2023. [Online]. Available: \url{http://arxiv.org/abs/2303.18223}
\BIBentrySTDinterwordspacing

\bibitem{zan-etal-2023-large}
\BIBentryALTinterwordspacing
D.~Zan, B.~Chen, F.~Zhang, D.~Lu, B.~Wu, B.~Guan, W.~Yongji, and J.-G. Lou, ``Large language models meet {NL}2{C}ode: A survey,'' in \emph{Proceedings of the 61st Annual Meeting of the Association for Computational Linguistics (Volume 1: Long Papers)}, A.~Rogers, J.~Boyd-Graber, and N.~Okazaki, Eds.\hskip 1em plus 0.5em minus 0.4em\relax Toronto, Canada: Association for Computational Linguistics, Jul. 2023, pp. 7443--7464. [Online]. Available: \url{https://aclanthology.org/2023.acl-long.411}
\BIBentrySTDinterwordspacing

\bibitem{Beller}
\BIBentryALTinterwordspacing
M.~Beller, A.~Bacchelli, A.~Zaidman, and E.~Juergens, ``Modern code reviews in open-source projects: which problems do they fix?'' in \emph{Proceedings of the 11th Working Conference on Mining Software Repositories}, ser. MSR 2014.\hskip 1em plus 0.5em minus 0.4em\relax New York, NY, USA: Association for Computing Machinery, 2014, p. 202–211. [Online]. Available: \url{https://doi.org/10.1145/2597073.2597082}
\BIBentrySTDinterwordspacing

\bibitem{nam2024}
\BIBentryALTinterwordspacing
D.~Nam, A.~Macvean, V.~Hellendoorn, B.~Vasilescu, and B.~Myers, ``Using an llm to help with code understanding,'' in \emph{ICSE '24: Proceedings of the IEEE/ACM 46th International Conference on Software Engineering}.\hskip 1em plus 0.5em minus 0.4em\relax New York, NY, USA: Association for Computing Machinery, 2024. [Online]. Available: \url{https://doi.org/10.1145/3597503.3639187}
\BIBentrySTDinterwordspacing

\bibitem{anagnostidis2024}
\BIBentryALTinterwordspacing
S.~Anagnostidis and J.~Bulian, ``How susceptible are llms to influence in prompts?'' in \emph{First Conference on Language Modeling}, 2024. [Online]. Available: \url{https://openreview.net/forum?id=y7JnjDcIQa}
\BIBentrySTDinterwordspacing

\bibitem{ebert2021exploratory}
F.~Ebert, F.~Castor, N.~Novielli, and A.~Serebrenik, ``An exploratory study on confusion in code reviews,'' \emph{Empirical Software Engineering}, vol.~26, pp. 1--48, 2021.

\bibitem{chouchen2021anti}
M.~Chouchen, A.~Ouni, R.~G. Kula, D.~Wang, P.~Thongtanunam, M.~W. Mkaouer, and K.~Matsumoto, ``Anti-patterns in modern code review: Symptoms and prevalence,'' in \emph{Proceedings of the IEEE international conference on software analysis, evolution and reengineering}.\hskip 1em plus 0.5em minus 0.4em\relax IEEE, 2021, pp. 531--535, acceptance rate: 46

\bibitem{Brown-etal-2020-few-shot}
\BIBentryALTinterwordspacing
T.~Brown, B.~Mann, N.~Ryder, M.~Subbiah, J.~D. Kaplan, P.~Dhariwal, A.~Neelakantan, P.~Shyam, G.~Sastry, A.~Askell, S.~Agarwal, A.~Herbert-Voss, G.~Krueger, T.~Henighan, R.~Child, A.~Ramesh, D.~Ziegler, J.~Wu, C.~Winter, C.~Hesse, M.~Chen, E.~Sigler, M.~Litwin, S.~Gray, B.~Chess, J.~Clark, C.~Berner, S.~McCandlish, A.~Radford, I.~Sutskever, and D.~Amodei, ``Language models are few-shot learners,'' in \emph{Advances in Neural Information Processing Systems}, H.~Larochelle, M.~Ranzato, R.~Hadsell, M.~Balcan, and H.~Lin, Eds., vol.~33.\hskip 1em plus 0.5em minus 0.4em\relax Curran Associates, Inc., 2020, pp. 1877--1901. [Online]. Available: \url{https://proceedings.neurips.cc/paper_files/paper/2020/file/1457c0d6bfcb4967418bfb8ac142f64a-Paper.pdf}
\BIBentrySTDinterwordspacing

\bibitem{Hou-etal-2024}
\BIBentryALTinterwordspacing
X.~Hou, Y.~Zhao, Y.~Liu, Z.~Yang, K.~Wang, L.~Li, X.~Luo, D.~Lo, J.~Grundy, and H.~Wang, ``Large language models for software engineering: A systematic literature review,'' \emph{ACM Trans. Softw. Eng. Methodol.}, Sep. 2024. [Online]. Available: \url{https://doi.org/10.1145/3695988}
\BIBentrySTDinterwordspacing

\bibitem{mu-etal-2024-navigating}
\BIBentryALTinterwordspacing
Y.~Mu, B.~P. Wu, W.~Thorne, A.~Robinson, N.~Aletras, C.~Scarton, K.~Bontcheva, and X.~Song, ``Navigating prompt complexity for zero-shot classification: A study of large language models in computational social science,'' in \emph{Proceedings of the 2024 Joint International Conference on Computational Linguistics, Language Resources and Evaluation (LREC-COLING 2024)}, N.~Calzolari, M.-Y. Kan, V.~Hoste, A.~Lenci, S.~Sakti, and N.~Xue, Eds.\hskip 1em plus 0.5em minus 0.4em\relax Torino, Italia: ELRA and ICCL, May 2024, pp. 12\,074--12\,086. [Online]. Available: \url{https://aclanthology.org/2024.lrec-main.1055}
\BIBentrySTDinterwordspacing

\bibitem{robinson2023leveraging}
\BIBentryALTinterwordspacing
J.~Robinson and D.~Wingate, ``Leveraging large language models for multiple choice question answering,'' in \emph{The Eleventh International Conference on Learning Representations}, 2023. [Online]. Available: \url{https://openreview.net/forum?id=yKbprarjc5B}
\BIBentrySTDinterwordspacing

\bibitem{our_replication_package}
``Replication package,'' \url{https://zenodo.org/records/15003074}.

\bibitem{qwen2models}
\BIBentryALTinterwordspacing
B.~H. An~Yang, Baosong~Yang \emph{et~al.}, ``Qwen2 technical report,'' 2024. [Online]. Available: \url{https://arxiv.org/abs/2407.10671}
\BIBentrySTDinterwordspacing

\bibitem{llama3herdmodels}
\BIBentryALTinterwordspacing
A.~P. Abhimanyu~Dubey, Abhinav~Jauhri \emph{et~al.}, ``The llama 3 herd of models,'' 2024. [Online]. Available: \url{https://arxiv.org/abs/2407.21783}
\BIBentrySTDinterwordspacing

\bibitem{Oleksii2016}
O.~Kononenko, O.~Baysal, and M.~W. Godfrey, ``Code review quality: how developers see it,'' in \emph{Proceedings of the 38th International Conference on Software Engineering}, ser. ICSE '16, 2016, p. 1028–1038.

\bibitem{Li2022CodeReviewer}
Z.~Li, S.~Lu, D.~Guo, N.~Duan, S.~Jannu, G.~Jenks, D.~Majumder, J.~Green, A.~Svyatkovskiy, S.~Fu, and N.~Sundaresan, ``Automating code review activities by large-scale pre-training,'' in \emph{Proceedings of ESEC/FSE}, 2022, p. 1035–1047.

\bibitem{Tufano2022PretrainedModels}
R.~Tufano, S.~Masiero, A.~Mastropaolo, L.~Pascarella, D.~Poshyvanyk, and G.~Bavota, ``Using pre-trained models to boost code review automation,'' in \emph{Proceedings of ICSE}, 2022, p. 2291–2302.

\bibitem{Lin2024}
H.~Y. Lin, P.~Thongtanunam, C.~Treude, and W.~Charoenwet, ``Improving automated code reviews: Learning from experience,'' in \emph{Proceedings of the IEEE/ACM International Conference on Mining Software Repositories}, 2024, p. to appear.

\bibitem{tufano2019learning}
M.~Tufano, J.~Pantiuchina, C.~Watson, G.~Bavota, and D.~Poshyvanyk, ``On learning meaningful code changes via neural machine translation,'' in \emph{Proceedings of ICSE}, 2019, p. 25–36.

\bibitem{thongtanunam2022autotransform}
P.~Thongtanunam, C.~Pornprasit, and C.~Tantithamthavorn, ``Autotransform: Automated code transformation to support modern code review process,'' in \emph{Proceedings of the IEEE/ACM International Conference on Software Engineering}, 2022, pp. 237--248.

\bibitem{pornprasit2023d}
C.~Pornprasit, C.~Tantithamthavorn, P.~Thongtanunam, and C.~Chen, ``D-act: Towards diff-aware code transformation for code review under a time-wise evaluation,'' in \emph{Proceedings of the IEEE International Conference on Software Analysis, Evolution and Reengineering}.\hskip 1em plus 0.5em minus 0.4em\relax IEEE, 2023, pp. 296--307.

\bibitem{Jason-etal-Chain}
\BIBentryALTinterwordspacing
J.~Wei, X.~Wang, D.~Schuurmans, M.~Bosma, b.~ichter, F.~Xia, E.~Chi, Q.~V. Le, and D.~Zhou, ``Chain-of-thought prompting elicits reasoning in large language models,'' in \emph{Advances in Neural Information Processing Systems}, S.~Koyejo, S.~Mohamed, A.~Agarwal, D.~Belgrave, K.~Cho, and A.~Oh, Eds., vol.~35.\hskip 1em plus 0.5em minus 0.4em\relax Curran Associates, Inc., 2022, pp. 24\,824--24\,837. [Online]. Available: \url{https://proceedings.neurips.cc/paper_files/paper/2022/file/9d5609613524ecf4f15af0f7b31abca4-Paper-Conference.pdf}
\BIBentrySTDinterwordspacing

\bibitem{pavlovic-poesio-2024-effectiveness}
\BIBentryALTinterwordspacing
M.~Pavlovic and M.~Poesio, ``The effectiveness of {LLM}s as annotators: A comparative overview and empirical analysis of direct representation,'' in \emph{Proceedings of the 3rd Workshop on Perspectivist Approaches to NLP (NLPerspectives) @ LREC-COLING 2024}, G.~Abercrombie, V.~Basile, D.~Bernadi, S.~Dudy, S.~Frenda, L.~Havens, and S.~Tonelli, Eds.\hskip 1em plus 0.5em minus 0.4em\relax Torino, Italia: ELRA and ICCL, May 2024, pp. 100--110. [Online]. Available: \url{https://aclanthology.org/2024.nlperspectives-1.11}
\BIBentrySTDinterwordspacing

\bibitem{pangakis-wolken-2024-knowledge}
\BIBentryALTinterwordspacing
N.~Pangakis and S.~Wolken, ``Knowledge distillation in automated annotation: Supervised text classification with {LLM}-generated training labels,'' in \emph{Proceedings of the Sixth Workshop on Natural Language Processing and Computational Social Science (NLP+CSS 2024)}, D.~Card, A.~Field, D.~Hovy, and K.~Keith, Eds.\hskip 1em plus 0.5em minus 0.4em\relax Mexico City, Mexico: Association for Computational Linguistics, Jun. 2024, pp. 113--131. [Online]. Available: \url{https://aclanthology.org/2024.nlpcss-1.9}
\BIBentrySTDinterwordspacing

\end{thebibliography}

\end{document}